\documentclass[12pt,aps,prd,tightenlines,preprintnumbers,%
            superscriptaddress,showpacs,nofootinbib]{revtex4}
\newcommand{\PRE}[1]{{#1}} 

\usepackage{bm} \usepackage{epsfig}

\newcommand{\fb}{\text{fb}} 
\newcommand{\ifb}{\text{fb}^{-1}}

\newcommand{\gev}{\text{GeV}} 
\newcommand{\tev}{\text{TeV}}

\newcommand{\s}{\text{s}}

\newcommand{\eqref}[1]{Eq.~(\ref{#1})}
\newcommand{\eqsref}[2]{Eqs.~(\ref{#1}) and (\ref{#2})}

\newcommand{\bold}[1]{{\text{\normalsize\bm{$#1$}}}}

\begin{document}

\preprint{UCI-TR-2005-26}  
\preprint{MIT-CTP-3629}  
\preprint{hep-ph/0507032}

\title{
\PRE{\vspace*{1.5in}}
Advantages and Distinguishing Features of \\
Focus Point Supersymmetry
\PRE{\vspace*{0.3in}}
}

\author{Jonathan L.~Feng}
\affiliation{Department of Physics and Astronomy, University of
California, Irvine, CA 92697, USA
\PRE{\vspace*{.3in}}
}

\author{Frank Wilczek%
\PRE{\vspace*{.2in}}
} 
\affiliation{Center for Theoretical Physics, Department of Physics,
Massachusetts Institute of Technology, Cambridge, Massachusetts 02139,
USA
\PRE{\vspace*{.3in}}
}


\begin{abstract} \PRE{\vspace*{.3in}} 
Diverse experimental constraints now motivate models of supersymmetry
breaking in which some superpartners have masses well above the weak
scale.  Three alternatives are focus point supersymmetry and inverted
hierarchy models, which embody a naturalness constraint, and the more
recent framework of split supersymmetry, which relaxes that
constraint.  Many aspects of their phenomenology are very similar.
They can be distinguished, however, through detailed study of
superoblique parameters, the Higgs potential and other observables.
\end{abstract}


\maketitle

\PRE{\newpage}

\section{Introduction}

The standard model of particle physics is fine-tuned.  Quantum
corrections to the scalar Higgs boson mass$^2$ are quadratically
divergent, so that a natural estimate of their magnitude is $\alpha
M^2$, where $M$ is a cutoff mass. If we associate the cutoff with
unification scale or Planck scale physics, we find that the quantum
corrections are much larger than the desired net result.  This blemish
has been a prime motivation for proposing supersymmetric extensions to
the standard model.  In models with low-energy supersymmetry,
naturalness can be restored by having superpartners with approximately
weak-scale masses~\cite{Maiani}.  Low-energy supersymmetry facilitates
several other theoretically desirable ideas, including, very notably,
quantitatively accurate unification of gauge
couplings~\cite{Dimopoulos:1981yj}.  It also provides an excellent
dark matter candidate~\cite{Goldberg:1983nd}.

Unfortunately, straightforward breaking of supersymmetry at the weak
scale also opens the door to various difficulties.  Together with many
new particles it introduces many new possibilities for couplings,
which generically induce unacceptable violations of observed
approximate symmetries.  Conservation of $R$-parity removes the most
severe of these difficulties, but significant challenges remain.
Superpartners are accompanied by many new flavor mixing angles and
$CP$-violating phases.  If those mixings and phases are of order
unity, then constraints on flavor-changing neutral currents and the
$\epsilon$ parameter require some superpartner masses to be at or
above $\sim 10~\tev$ and 100 TeV, respectively~\cite{Ciuchini:1998ix}.
If flavor mixing is suppressed, but $CP$-violating phases aren't, the
electron and neutron electric dipole moments still require some
superpartners to have masses above
$2~\tev$~\cite{Pospelov:2005pr,Olive:2005ru}.  Finally, bounds
arising from theoretical estimates of proton decay and the Higgs boson
mass are most easily obeyed if some superpartners have masses well
above the weak scale~\cite{Goto:1998qg,Ambrosanio:2001xb}.  While none
of these constraints is completely watertight, taken together they put
considerable pressure on models that attempt to keep all superpartner
masses close to the weak scale.

An alternative is to take the data at face value and explore the most
straightforward interpretation: that some superpartners are
superheavy, with masses well above the weak scale.  Here we briefly
compare and contrast conceptual frameworks for superheavy
supersymmetry: focus point
supersymmetry~\cite{Feng:1999hg,Feng:1999mn,Feng:2000gh,Feng:2000bp},
which is our primary emphasis, inverted hierarchy
models~\cite{Drees:1985jx,Dvali:1996rj}, and split
supersymmetry~\cite{Arkani-Hamed:2004fb,Giudice:2004tc}.
Operationally, below and even at LHC energies, they appear rather
similar, for in all, the central proposal is to allow squark and
slepton masses to be large, while keeping gaugino masses relatively
small.  Philosophically, however, they are quite different: focus
point supersymmetry retains naturalness of the weak scale as a guiding
principle and implements it through a dynamical mechanism, inverted
hierarchy models retain naturalness for the weak scale and implement
it by hypothesizing a specific family-dependent pattern of
supersymmetry breaking masses, while split supersymmetry explicitly
abandons naturalness.
  
Since the robust phenomenological and cosmological features of the
focus point and split supersymmetry frameworks, first examined in
detail in
Refs.~\cite{Feng:1999hg,Feng:1999mn,Feng:2000gh,Feng:2000bp}, are so
similar, refined measurements will be needed to decide between them.
We outline how measurements of superoblique parameters and other
practical observables can accomplish that task.  If we discover,
through the appearance of gauginos but not squarks and sleptons at the
Large Hadron Collider (LHC), that a structured form of supersymmetry
breaking holds in nature, it will be important to carry out such
measurements to elucidate the conceptual meaning of the discovery.

\section{Focus Point Supersymmetry}

Focus point supersymmetry is defined by the hypothesis that all
squarks and sleptons are superheavy, with masses at the TeV scale or
higher, while gauginos and Higgsinos remain at the weak scale, and the
hypothesis that the weak scale arises naturally.  There is tension
between these hypotheses, but no
contradiction~\cite{Feng:1999hg,Feng:1999mn}.  The naturalness
requirement, that the electroweak potential is insensitive to small
relative changes in the fundamental supersymmetry breaking parameters,
can either be met straightforwardly, by having all these parameters
small, or through focusing.  In the latter alternative,
renormalization group evolution focuses a large range of initial
values, defined by the fundamental parameters at the unification
scale, into a relatively small range of effective values for the
phenomenologically relevant parameters at the weak scale.

In practice, insensitivity of the weak scale to variations in the
fundamental parameters is largely guaranteed if focusing occurs for
the up-type Higgs boson mass.  It will occur if the soft scalar masses
at the unification scale are in the ratio~\cite{Feng:1999mn}
\begin{equation}
( m_{H_u}^2, m_{\tilde{t}_R}^2, m_{\tilde{t}_L}^2 ) \propto 
(1, 1+x, 1-x)
\label{moderatetb}
\end{equation}
for moderate values of $\tan\beta$, and 
\begin{equation}
( m_{H_u}^2, m_{\tilde{t}_R}^2, m_{\tilde{t}_L}^2,
m_{\tilde{b}_R}^2, m_{H_d}^2 ) \propto 
(1, 1+x, 1-x, 1+x-x', 1+x' ) 
\label{largetb}
\end{equation}
for large values of $\tan\beta$, where $x$ and $x'$ are arbitrary
constants.  A universal scalar mass obviously satisfies both
\eqsref{moderatetb}{largetb}, but in principle more general
possibilities are allowed.  Given \eqref{moderatetb} or
\eqref{largetb}, focusing occurs for any weak-scale gaugino masses and
$A$-parameters, any moderate or large value of $\tan\beta$, and any
top quark mass within existing experimental bounds.  Note that
focusing makes the weak scale insensitive to variations in parameters
introduced to explain the weak scale, the supersymmetry breaking
parameters, but not to variations in other parameters, such as the top
quark Yukawa coupling.  Of course, the fact that the measured top
quark mass is compatible with focusing for simple boundary conditions
is tantalizing, if preliminary, quantitative evidence for focus point
supersymmetry.

Focus point supersymmetry has been studied in great detail for the
specific case of minimal supergravity.  For top quark mass $m_t = 174\
(178)~\gev$, the region in which all phenomenological constraints are
satisfied and relic neutralino dark matter has the observed density is
at $m_0 \sim 3 \ (8)~\tev$~\cite{Feng:1999mn,Paige:2003mg}.  Such
superheavy squarks and sleptons sufficiently suppress one-loop
contributions to the electron and neutron electric dipole moments even
for ${\cal O}(1)$ phases.  Two-loop effects are dominant and might be
within experimental reach in the near future~\cite{Chang:1998uc}.  The
high sfermion masses, together with additional suppression from squark
and slepton degeneracy as occurs in unified focus point models,
comfortably solve all problems with flavor-violation and
flavor-violating CP-violation~\cite{Feng:2000bp}.  Of course, given
the Tevatron Run I average top mass of $m_t = 178.0 \pm
4.3~\gev$~\cite{Azzi:2004rc} and the most recent average including
preliminary Run II results of $m_t = 174.3 \pm
3.4~\gev$~\cite{Group:2005cg}, values of $m_t$ higher than $178~\gev$
are still well within current constraints.  For such top masses, the
focus point region moves to values of $m_0 \agt 10~\tev$.  In this
regime the heaviness of squarks and sleptons can remove all the flavor
and CP problems associated with low-energy supersymmetry without the
need for flavor degeneracy or additional assumptions.

A broad variety of phenomenological implications and virtues of the
focus point spectrum has been explored more generally in
Refs.~\cite{Feng:1999hg,Feng:1999mn,Feng:2000gh,Feng:2000bp}:
\begin{itemize}
\item A noteworthy feature is that radiative corrections to the
predicted value of the Higgs boson mass arising from loops containing
heavy top and bottom squarks can raise the Higgs boson mass well above
current bounds~\cite{Feng:2000bp}.  This feature does not occur for
inverted hierarchy models~\cite{Drees:1985jx,Dvali:1996rj}, in which
the light fermions have superheavy partners, while the heavy fermions
have light (weak-scale) superpartners.  Like focus point
supersymmetry, inverted hierarchy models resolve many of the
phenomenological difficulties generically associated with low-energy
supersymmetry without sacrificing naturalness, because experimental
constraints are stringent only for observables involving the first two
generations, while naturalness constraints are stringent only for
fields with large couplings to the Higgs sector~\cite{Drees:1985jx}.

\item Gauge unified focus point models naturally obey constraints on
proton decay as well~\cite{Feng:2000bp}.  Viewed in isolation,
suppression of proton decay does not pose a critical problem: the
dangerous processes involve virtual exchange of both standard model
superpartners and unification-scale particles, especially the color
triplet Higgs superpartners, and they can always be satisfied by
raising the masses of the latter.  But if we want to maintain the
impressive quantitative success of the unification of couplings, which
is a major motivation for low-energy supersymmetry, then obtaining
sufficient suppression of proton decay is
problematic~\cite{Bajc:2002pg}.  Coupling constant unification
constrains unification-scale threshold effects, which in simple
unification models implies upper bounds on GUT-scale masses.  With
superheavy squarks and sleptons, this difficulty is resolved, and one
is left with viable (and interesting!) expectations for proton decay.

\item In focus point models the lightest supersymmetric particle (LSP)
is a neutralino that provides a dark matter candidate with excellent
prospects for detection~\cite{Feng:2000gh}. In this
context, the neutralino cannot be pure Bino, because in that case it
annihilates through $\tilde{B} \tilde{B} \to f \bar{f}$ with a
$t$-channel sfermion $\tilde{f}$, and these processes become
inefficient for $m_{\tilde{f}}$ in the multi-TeV range or above,
leading to an overabundant relic density.  For neutralinos with
significant Wino or Higgsino component, however, $\chi \chi \to WW$
and $\chi \chi \to ZZ$ become efficient, and the LSP's relic density
is naturally in the desired range.  For similar reasons, mixed
neutralinos give rise to relatively large direct and indirect
detection rates.
 \end{itemize}

\section{Abandoning Naturalness?}

The confluence of the existing failure to explain the anomalously
small value of the cosmological term in a natural way, the suggestion
from inflationary scenarios that on ultra-ultra-large scales the
Universe might be drastically inhomogeneous, and the longstanding
indications that consistent solutions of the equations of string
theory provide a plethora of candidate macroscopic
universes~\cite{Bousso:2000xa} have rekindled interest in the
possibility that selection effects (random or anthropic) play a more
central role, and the program of explanation through symmetry and
naturalness a less central role, than traditionally has been assumed
in theoretical physics.  While it is certainly logically possible that
one will be driven in that direction, we feel that it is a wise
methodological principle to attempt to maintain the tightest available
explanatory framework until forced to abandon it.  Moreover, in
several specific instances, including the unification of couplings,
the smallness of the $\theta$ term in QCD, and the extremely long
lifetime of the proton, it is difficult to conceive of plausible
selection effects that could supplant symmetry as an explanation of
the observed phenomena.

The central proposal of split supersymmetry is to drop any direct
connection between low-energy supersymmetry and the solution of the
weak scale hierarchy problem
~\cite{Arkani-Hamed:2004fb,Giudice:2004tc}.  On the face of it, that
idea would suggest that all superpartners acquire unification or
Planck-scale masses, if indeed one has supersymmetry at all.  To
preserve desirable features of low-energy supersymmetry, i.e.,
quantitative unification of couplings and the existence of a good dark
matter candidate, however, additional residual symmetries (and
fine-tunings, see below) are postulated to ensure that there are
gauginos and Higgsinos with weak-scale masses.  Thus,
phenomenologically, split supersymmetry is very similar to focus point
supersymmetry, but one no longer requires \eqref{moderatetb} or
\eqref{largetb}, and the squark and slepton masses are allowed to
become arbitrarily large.

Are the distinctions testable? The answer is not immediately obvious,
because those distinctions lie in the masses of the superheavy
superpartners, which are beyond the reach of currently planned
colliders and largely decouple from low energy observables.

\section{Tests of Naturalness}

One might hope to distinguish focus point and split supersymmetry by
finding evidence for extremely large squark and slepton masses.
Extremely heavy sfermions lead, through radiative corrections, to
large Higgs boson masses, for example.  An even more striking
prediction is that, for extremely heavy squarks, gluinos become
long-lived, with lifetime~\cite{Kilian:2004uj}
\begin{equation}
\tau_{\tilde{g}} \sim 10^{-12}~\s \ 
\left[ \frac{m_{\tilde{q}}}{10^6~\gev} \right]^4
\left[ \frac{1~\tev}{m_{\tilde{g}}} \right]^5 \ .
\label{longlife}
\end{equation}
Long-lived, weak-scale gluinos have been studied in
Refs.~\cite{Farrar:1978xj}.  They arise in theories with weak-scale
supersymmetry breaking where the gluino is the LSP or decays only to a
gravitino LSP.  Those studies motivated discussions of the
accompanying collider phenomenology and appropriate triggers long
before the proposal of split supersymmetry.  Nevertheless, coexistence
of long-lived gluinos with lighter neutralinos and charginos could
provide an unambiguous signal of superheavy sfermions.

Unfortunately, for \eqref{longlife} to yield a practically detectable
lifetime, sfermion masses probably must exceed $10^6~\gev$.  Such
large masses pose a significant challenge, because Weyl
anomaly-mediated contributions~\cite{Randall:1998uk,Giudice:1998xp}
require gaugino/Higgsino masses to be suppressed relative to sfermion
masses by no more than a factor of $\sim g^2 / (16 \pi^2)$.  If such
contributions are present, then, the natural range for the superheavy
sfermion masses is constrained to be at or below $10^5~\gev$.  Of
course, given the few guiding principles in split supersymmetry,
there is no requirement that anomaly-mediated contributions be present
at the expected order of magnitude.

Both split supersymmetry and focus point supersymmetry can accommodate
superheavy superpartner masses in the $10^4$ to $10^5~\gev$ range.  As
noted above, the focus point mechanism preserves naturalness for $m_t
= 178~\gev$ for scalar masses $\sim 10~\tev$ and weak-scale gauginos
and Higgsinos.  However, the preferred sfermion mass range depends on
the top quark mass and increases rapidly for larger $m_t$.  A careful
analysis of renormalization group equations and electroweak symmetry
breaking is required to determine the exact relation.  However, given
the currently favored range of top quark masses, large sfermion masses
above 10 TeV are certainly a possibility, and the mere presence of
sfermion masses in this range cannot be used to distinguish between
natural and fine-tuned theories.

A far more incisive method for differentiating superheavy particle
spectra is through superoblique parameters~\cite{Cheng:1997sq}.
Superoblique parameters measure splittings between dimensionless
couplings and their supersymmetric analogues.  Exact supersymmetry
demands equality of these couplings, but split supermultiplets
introduce corrections~\cite{Chankowski:1989du,Hikasa:1996bw}.  As with
their electroweak analogues, the oblique
corrections~\cite{Peskin:1990zt}, superoblique corrections are
non-decoupling: they become {\em large} for highly split
supermultiplets. They can be determined by precise measurements of the
properties of light superpartners, which are kinematically accessible
in both focus point and split supersymmetry frameworks.  These
properties imply that superoblique parameters are likely to play an
essential role in the experimental exploration of any supersymmetric
theory in which some superpartners are beyond direct detection.

The full set of possible superoblique parameters has been
cataloged~\cite{Cheng:1997vy}, and their measurement at colliders has
been explored in detail in several studies~\cite{Cheng:1997vy,%
Feng:1995zd,Nojiri:1996fp,%
Nojiri:1997ma,Katz:1998br,Kiyoura:1998yt,Mahanta:1999hx}.\footnote{The
super-oblique parameters have also recently been discussed again in
the context of split supersymmetry, for example in
Ref.~\protect\cite{Arkani-Hamed:2004fb}, where a subset of them have
been reparametrized and denoted $\kappa$.}  In the leading logarithm
approximation, the superoblique parameters are
\begin{equation}
\tilde{U}_i \equiv \frac{h_i}{g_i} - 1 
\approx \frac{g_i^2}{16 \pi^2} \left( b_{g_i} - b_{h_i} \right) 
\times \ln R \ , 
\end{equation}
where $i = 1,2,3$ denotes the gauge group U(1), SU(2), or SU(3), $g_i$
is the standard model gauge coupling, $h_i$ its supersymmetric
analogue, and $R$ is the ratio between the effective superheavy
superpartner mass scale and the weak scale. The coefficients $b_{g_i}$
and $b_{h_i}$ are the one-loop beta function coefficients for $g_i$
and $h_i$ for the effective theory between the superheavy and weak
scales; $b_{g_i} - b_{h_i}$ is therefore the contribution from
standard model particles whose superpartners are superheavy.  For
focus point supersymmetry and split supersymmetry in which all
sfermions are superheavy, the superheavy particles are in complete
multiplets of SU(5), and so $b_{g_i} - b_{h_i}$ is independent of $i$.
Numerically, $b_{g_i} - b_{h_i} = 4$, and
\begin{eqnarray}
\tilde{U}_1 &\approx& 1.2\% \ \log_{10} R \\
\tilde{U}_2 &\approx& 2.5\% \ \log_{10} R \\
\tilde{U}_3 &\approx& 8.3\% \ \log_{10} R  \ .
\end{eqnarray}

In focus point supersymmetry and split supersymmetry, the superoblique
parameters can be measured in a number of ways.  As an example,
consider the chargino mass matrix
\begin{equation}
\bold{M}_{\chi^{\pm}}
= \left( \begin{array}{cc}
 M_2                    &\frac{1}{\sqrt{2}} h_2 v \sin\beta \\
\frac{1}{\sqrt{2}} h_2 v \cos\beta   &\mu    \end{array} \right) \ .
\end{equation}
In the limit of exact supersymmetry, the $Whh$ and $\tilde{W}
\tilde{h} h$ couplings are identical, and so $h_2$ is equal to $g_2$,
the SU(2) gauge coupling constant.  Superheavy superpartners break
this degeneracy, and predict a non-vanishing superoblique parameter
$\tilde{U}_2$.  Dark matter constraints require significant mixing in
the neutralino and chargino sectors, and so it is likely that both
charginos and all four neutralinos will be produced at the Large
Hadron Collider and the International Linear Collider.  

The possibility of measuring superoblique parameters at the
International Linear Collider in scenarios with mixed charginos and
neutralinos has been discussed in
Refs.~\cite{Feng:1995zd,Cheng:1997vy,Kiyoura:1998yt}.  Supersymmetric
parameters may be constrained by measuring chargino and neutralino
masses and bounding the polarized cross sections for chargino and
neutralino pair production.  The sensitivity to the superheavy mass
scale entering through the dependence of the chargino mass matrix on
$\tilde{U}_2$ may be quite large.  For example, in the mixed scenario
studied in Ref.~\cite{Kiyoura:1998yt}, the cross section $\sigma_R =
\sigma ( e^-_R e^+ \to \chi^+_1 \chi^-_1)$ varies from $\sim 50~\fb$
to 62 fb as the superheavy scalar mass scale varies from 1 to 10 TeV.
Given an integrated luminosity of $50~\ifb$, the statistical
uncertainty in $\sigma_R$ is $\sim 2 \%$, corresponding to an
uncertainty in the superheavy mass scale of $\Delta \log_{10} R \sim
0.1$.  Of course, this precision will be compromised by systematic
experimental uncertainties and uncertainties in other supersymmetry
parameters.  The size of these effects depends on the underlying
supersymmetry scenario realized in nature, the final properties of the
International Linear Collider, and the success with which other
experiments may be used to constrain supersymmetry parameters, such as
$\tan\beta$.  Nevertheless, barring the possibility that these effects
completely degrade the statistical precision, constraints on the
superheavy superpartner mass scale to within an order of magnitude
($\Delta \log_{10} R \sim 1$) appear possible.

Fine structure within the superheavy superpartner mass spectrum may be
constrained by precise measurements of branching fractions mediated by
virtual superheavy superpartners.  The branching fractions
$B(\tilde{g} \to q_R \bar{q}_R \chi )$ and $B(\tilde{g} \to q_L
\bar{q}_L \chi )$ are sensitive to the fourth powers of
$m_{\tilde{q}_R}$ and $m_{\tilde{q}_L}$, respectively. For the cases
of greatest interest here, where $q = t, b$, these branching fractions
with polarized final states can be distinguished through the energy
distributions of $q$ decay products.  Splittings in the superheavy
spectrum also result in different effective $R$ parameters for the
different superoblique parameters, and so additional rough constraints
on fine structure can also be obtained if the superoblique parameters
can be measured in more than one way.  Finally, $m_{H_u}^2$ and
$m_{H_d}^2$ can be determined by precise measurements of $\mu$,
$\tan\beta$, and other parameters entering the Higgs potential.

These weak scale parameters can then be extrapolated to high scales to
determine the fundamental soft supersymmetry-breaking parameters.
This program is challenging.  However, if superheavy masses above 100
TeV are realized in nature, even rough constraints on the superheavy
mass scale will likely provide evidence for fine-tuning or,
alternatively, motivate focusing or other mechanisms different from
those discussed so far.  On the other hand, consistency with
superheavy mass scales below 100 TeV and with the predictions of
\eqsref{moderatetb}{largetb} would constitute striking evidence for
focus point supersymmetry and naturalness.  It would further motivate
mechanisms of supersymmetry breaking that explain
\eqsref{moderatetb}{largetb}, providing essential guidance for the
next step to more fundamental theories.

\section*{Acknowledgments}

The work of JLF is supported in part by NSF CAREER grant
No.~PHY-0239817, NASA Grant No.~NNG05GG44G, and the Alfred P.~Sloan
Foundation.  The work of FW is supported in part by funds provided by
the U.S. Department of Energy under cooperative research agreement
DE-FC02-94ER40818.



\end{document}